\documentclass[a4paper,12pt]{iopart}
\usepackage[utf8x]{inputenc}

\usepackage{ifpdf}

\ifpdf
   \usepackage[pdftex]{graphicx}
   \DeclareGraphicsExtensions{.pdf}
   \usepackage{thumbpdf}      
\else
   \usepackage[dvips]{graphicx}
   \DeclareGraphicsExtensions{.eps}
\fi

\usepackage{listings}

\usepackage{graphicx}

\usepackage{cite}
\usepackage{amscd}
\usepackage{subfigure}

\newcommand{\ri}{{ \rm i }}
\newcommand{\re}{{ \rm e }}
\newcommand{\rd}{{ \rm d }}

\newcommand{\be}{\begin{equation}}
\newcommand{\ee}{\end{equation}}

\renewcommand{\vec}[1]{\mathbf{#1}}

\newcommand{\text}[1]{{\mbox{#1}}}

\usepackage{color}
\definecolor{blau}{rgb}{0,0,1}
\definecolor{gruen}{rgb}{0,1,0}
\definecolor{rot}{rgb}{1,0,0}
\definecolor{magenta}{rgb}{1,0,1}

\begin{document}

\title{Calculating resonance positions and widths using the Siegert approximation method}
\author{Kevin Rapedius}

\address{Center for Nonlinear Phenomena and Complex Systems, Universit\'e Libre de Bruxelles, Code Postal 231, Campus Plaine,
B-1050 Brussels, Belgium}
\ead{kevin.rapedius@ulb.ac.be}

\begin{abstract}
Here we present complex resonance states (or Siegert states), that describe the tunneling decay of a trapped quantum particle, from an intuitive point of view which naturally leads to the 
easily applicable Siegert approximation method that can be used for analytical and numerical calculations of complex resonances of both the linear and nonlinear Schr\"odinger equation. Our approach thus complements other treatments of the subject 
that mostly focus on methods based on continuation in the complex plane or on semiclassical approximations. 

\end{abstract}

\pacs{03.65.-w, 03.65.Xp, 03.75.Lm, 73.40.Gk, 01.40.-d}

\maketitle

\section{Introduction}
The tunneling decay of particles trapped in an external potential $V(\vec{x})$ is a problem often encountered in nuclear, atomic and molecular physics. The most famous example is arguably the 
alpha decay of nuclei for which Gamow calculated the decay rates semiclassically \cite{Gamo28}. 

For a particle with mass $m$ such a decay can be described by resonance eigenstates $\psi_S$ of the Schr\"odinger equation
\be
  \left( - \frac{\hbar^2}{2m} \nabla^2 + V(\vec{x}) \right) \psi_S=\cal{E} \psi_S
\label{SE_comp}
\ee
with complex eigenenergy ${\cal E}=E_S-\ri \Gamma/2$, often refered to as Siegert resonances, where the wavefunction satisfies outgoing wave (or Siegert) boundary conditions \cite{Sieg39}, i.~e. the wavefunction 
is given by an outgoing plane wave for $|\vec{x}| \rightarrow \pm \infty$ (cf.~ equation (\ref{BC_S})). 
The imaginary part is also refered to as the decay rate since it leads to an exponential decay of the wavefunction,
\be
 \fl \quad \quad \quad \psi_S(\vec{x},t)=\exp(\ri {\cal E }t/\hbar) \psi_S(\vec{x},0)=\exp(-\Gamma t/(2\hbar)) \exp(\ri E_S t/\hbar) \psi_S(\vec{x},0) \,.
\label{wf_decay}
\ee

On the other hand, the tunneling decay of trapped particles is closely related to the problem of scattering of particles off the same potential since quantities like the transmission coefficient
 $|T(E)|^2$ or the scattering cross section show characteristic peaks near the resonance energies ${\cal E}$, which can be described by a Lorentz or Breit-Wigner profile
\be
  |T(E)|^2=\frac{(\Gamma/2)^2}{(E-E_S)^2+(\Gamma/2)^2}
\ee
the width of which is given by the decay rate $\Gamma$ \cite{Sieg39}.

As mentioned above, the decay rates can be calculated in the manner of Gamow using semiclassical approximations. Though these approximations are straightforward and easy to use they 
are often not very precise, only providing an order of magnitude. On the other hand there are powerful complex-scaling based methods (see e.~g.~\cite{Mois98,Mois11}) (including related methods such as complex absorbing 
potentials) where the spatial coordinate is rotated to the complex plane, $\vec{x} \rightarrow \vec{x} \exp(\ri \theta)$ by some sufficiently large angle $\theta$ to make the resonance wavefunctions square 
integrable, enabling the use of the usual techniques for calculating ordinary bound states. These methods are precise and highly efficient yet quite sophisticated and, apart from rare exceptions,
only suited for numerical calculations. 
Alternative techniques, like, e.g., the stabilization method \cite{Mand93a} have both some advantages and disadvantages compared to complex scaling 
and are, however, not very intuitive and only aim at numerical applications. While there are a number of excellent texts (see, e.~g.~\cite{Mois98,Tayl72,Mois11}
and references therein) that discuss the mathematical aspects of the problem (e.~g.~analytical continuation of the wavefunction in the complex plane) and the computational methods mentioned above, a 
complementary treatment assuming a different point of view could prove valuable.

In this article we want to draw attention to the scarcely known Siegert approximation method for calculating complex resonances {\cite{Sieg39,08nlLorentz}} which is both intuitive and easy to implement but does 
not rely on semiclassical arguments. It yields good results for narrow resonances (i.e. $\Gamma/2 \ll E_S$, where the resonance energy $E_S$ is measured relative to the potential energy at $|\vec{x}| \rightarrow \infty$) 
and it can in some cases lead to closed form analytical approximations for the decay coefficient. 
Another advantage of this method is its straightforward applicability to resonances of the nonlinear Schr\"odinger equation that occur, e.~g., in the context of trapped Bose-Einstein 
condensates \cite{08nlLorentz,09ddshell} since it does not require properties like the linearity or analyticity of the differential equation. While direct complex scaling \cite{Schl06a,Schl06b} has 
succesfully been applied to nonlinear Siegert resonances it is much less efficient than in the linear case and requires substantial modifications. Complex absorbing potential methods, which have proven
 more efficient in this context \cite{Mois04a,10nlws}, also require considerable modifications compared to the linear case.   

In the derivation of the Siegert approximation method given here, complex resonances are reviewed from an alternative, rather intuitive point of view which complements the usual more mathematical
treatments of the problem by emphasizing some important aspects like the role of the continuity equation in this context as well as the similarities and differences between 
complex resonance states and so-called transmission resonances, i.~e.~scattering states corresponding to the maxima of the transmission coefficient (or scattering cross section respectively) 
for real eigenenergies. 

This paper is organized as follows: In section \ref{sec:Sieg} the Siegert approximation method is described, focussing on resonances of the one-dimensional linear Schr\"odinger equation for the 
sake of simplicity. In section \ref{sec:apps} the method is illustrated by means of several analytical and numerical applications. Finally, the main aspects of the 
article are summarized in section \ref{sec:summary}. \ref{app:num} contains a MATLAB code for calculating resonances.

\section{The Siegert approximation method}
\label{sec:Sieg}

For narrow resonances (i.e. $\Gamma/2 \ll E$), one can often neglect the decay rate $\Gamma$ in (\ref{SE_comp}) and thus obtain an approximate resonance wavefunction $\psi_{\rm approx}$ and an approximate real part $E_{\rm approx}$ 
of the resonance energy with very little effort.

For the sake of simplicity we first consider symmetric one-dimensional finite range potentials
\be
   V(x)=\left\{
                \begin{array}{cl}
		0 & |x|>a\\
		V_x(x) & |x| \le a
		\end{array}
        \right.
\ee
with finite range $a$ and $V(-x)=V(x)$. An example of such a potential is the double barrier shown in the right panel of figure \ref{fig:Siegert} (bold blue curve) which can safely be assumed to be approximately equal to 
zero for $|x| \ge 20$. Also shown is the square $|\psi(x)|^2$ of the most stable resonance wavefunction which is strongly localized between the potential maxima and inherits the symmetry of the 
double barrier potential.

Apart from the Siegert resonances $\psi_S$ with complex energies $E_S -\ri \Gamma/2$ obtained for outgoing wave boundary conditions
\be
    \psi_S'(\pm a)= \pm \ri k_S \psi_S(\pm a), \quad \quad k_S=\sqrt{2m(E_S-\ri \Gamma/2)}/\hbar
\label{BC_S}
\ee
(the prime denotes a derivative by $x$) these potentials also possesses so-called transmission (or unit) resonances $\psi_T(x)$ corresponding to real energies $E_T$ for which the potential is 
completely transparent, i.e. for the corresponding transmission coefficient we have
 $|T(E_T)|^2=1$. We will see further below, that for narrow resonances $\psi_T(x)$ and $E_T$ are good approximations to $\psi_S$ and $E_S$ respectively.

To obtain the transmission coefficient $|T|^2$ we consider transmission through the potential $V(x)$ which is characterized by the following boundary conditions for the scattering 
wavefunction $\psi(x)$: On the left hand side we have a superposition of an incoming and a reflected plane wave
\be
   \psi(x)=A \exp(\ri kx)+B \exp(-\ri kx), \quad x <-a
\ee 
where $k=\sqrt{2mE}/\hbar$ is the wavenumber corresponding to an energy $E$ of the incoming wave. On the right hand side we only have an outgoing wave,
\be
   \psi(x)=C \exp(\ri kx) , \quad x>a \,.
\ee  
Thus the transmission coefficient reads
\be
   |T|^2=|C/A|^2 \,.
\ee
We immediately see that for the transmission resonances with $|T|^2=1$ we have $|C|=|A|$ (and $B=0$), so that we can always achieve $C=A$ by multiplying the wavefunction with 
a constant phase factor. Thus the boundary conditions for the transmission resonces $\psi_T$ can be written as
\be
    \psi_T=A \exp(\ri k_T x), \quad |x|>a 
\ee
or equivalently as
\be
    \psi_T'(\pm a)=i k_T \psi_T(\pm a) \text { with } k_T=\sqrt{2m E_T}/\hbar \,.
\label{BC_T}
\ee
The symmetry of both the Siegert resonance wavefunction $|\psi_S(-x)|^2=|\psi_S(x)|^2$ and the transmission resonance wavefunction $|\psi_T(-x)|^2=|\psi_T(x)|^2$ imply that the derivatives of 
$|\psi_S|^2$ and $|\psi_T|^2$ must vanish at $x=0$. Thus the respective boundary conditions (\ref{BC_S}) and (\ref{BC_T}) can be recast in the form
\be
    (|\psi_\alpha(x)|^2)'\big|_{x=0}=0, \quad  \psi_\alpha'(a)=\ri k_\alpha \psi_\alpha(a), \quad \alpha \in \{S,T\}
\label{BC_Sym}
\ee
with $k_\alpha=\sqrt{2m {\cal E}_\alpha}/\hbar$, ${\cal E}_S=E_S -\ri \Gamma/2$, ${\cal E}_T=E_T$.

Therefore it is quite intuitive that for $\Gamma/2 \ll E_S$ we can make the approximations $E_S \approx E_T$ and 
\be
   \psi_S(x) \approx\left\{
                \begin{array}{cl}
		\psi_T(x)   & 0 <x \le a\\
	         \psi_T(-x) & -a \le x<0
		\end{array}
        \right. \,.
\label{psi_app}
\ee
 
This was shown rigorously by Siegert \cite{Sieg39}. Note that this approximate correspondance between complex energy Siegert resonances describing decay and real energy transmission resonances, which generally holds for narrow resonance widths, is 
in fact one of the main reasons for considering complex resonances in the context of scattering \cite{Sieg39}. Now we have approximations for the wavefunction and the real part of the eigenenergy but what about the imaginary part, 
i.e. the decay rate ?

To this end let us assume that the potential $V(x)$ has local maxima at $x= \pm b$ with $0< b < a$ (as, for example, the potential shown in figure (\ref{fig:Siegert}) ). Then the probability of finding the particle described by our Schr\"odinger equation 
(\ref{SE_comp}) 'inside' the potential well, i.e. in the region $ -b \le x \le +b$ between the potential maxima is given by the norm 
\be
   N=\int_{-b}^b |\psi_S(x)|^2 \rd x \approx 2 \int_{0}^b |\psi_T(x)|^2 \rd x\,
\ee
of the wavefunction inside the well.
The exponential decay behaviour of the resonance wavefunction given in equation (\ref{wf_decay}) implies that the norm $N$ decays according to
\be
    \partial_t N=-\frac{\Gamma}{\hbar}N
\ee
so that the decay rate can be written as
\be
    \Gamma=-\hbar \frac{\partial_t N}{N} \,.
  \label{gamma}
\ee
The time derivative $\partial_t N= \int_{-b}^b \partial_t |\psi_S(x,t)|^2 \rd x \approx 2 \int_{0}^b \partial_t |\psi_T(x,t)|^2 \rd x$ of the norm can be found by means of the continuity equation for 
the resonance wavefunction $\psi_S$ which reads
\be
   \partial_t \rho_S = - j_S'
\label{cont}
\ee
with $\rho_S(x,t)=|\psi_S(x,t)|^2$ and the probability current 
\be
   j_S= - \frac{\ri \hbar}{2m} \left(\psi_S^* \psi_S'-\psi_S \psi_S'^* \right) \,.
\ee
Integrating the continuity equation (\ref{cont}) from $x=-b$ to $x=b$ we obtain $\partial_t N= -(j_S(b)-j_S(-b))$. Equation (\ref{psi_app}) implies that the currents are approximately given by
$j_S(b) \approx j_T(b)$ and $j_S(-b) \approx -j_T(b)$. Furthermore, the current $j_T$ corresponding to the transmission resonance $\psi_T$ does not depend on the position $x$ and in particular $j_T(b)=j_T(a)$. 
For $x \ge a$ the transmission resonance wavefunction is given by a plane wave $\psi_T(x)=\psi_T(a) \exp(\ri k_T (x-a)$ so that the current at $x=a$ simply reads
$j_T(a)=\hbar k_T |\psi(a)|^2/m$. Thus the decay coefficient (\ref{gamma}) becomes
\be
    \Gamma \approx \hbar \frac{j_T(a)}{\int_{0}^b |\psi_T(x)|^2 \rd x} = \frac{\hbar^2 k_T}{m} \frac{|\psi_T(a)|^2}{\int_{0}^b |\psi_T(x)|^2 \rd x}\,.
  \label{Gamma_s}
\ee
We call (\ref{Gamma_s}) the Siegert formula. Note that it only depends on $E_T$ and $\psi_T$ so that the exact values $E_S$ and $\psi_S$ are not required.

In general, the Siegert approximation method consists of two steps: 
\begin{enumerate}
 \item \label{I1} Neglect at first the imaginary part $\Gamma/2$ (also called decay coefficient) of the complex resonance energy in the Schr\"odinger equation for the Siegert resonance wave function in order to obtain 
       approximations to both the wave function and the real part of the resonance energy. (In the symmetric barrier case discussed above this is done by calculating the energies and scattering states 
       for which the transmission probability of the corresponding scattering problem has a local maximum or, equivalently, directly use the symmetry of the problem expressed in the boundary conditions given 
       in equation (\ref{BC_Sym}).)
 \item \label{I2} Use the continuity equation to obtain a Siegert formula, i.~e.~an approximate expression for the decay coefficient $\Gamma$ (as given in equation (\ref{Gamma_s}) for the symmetric barrier case) which only 
       depends on the approximate energy and approximate wavefunction calculated in step (\ref{I1}).
\end{enumerate}

The Siegert approximation method yields good results for narrow resonances, i.~e.~whenever the imaginary part $\Gamma/2$ of the resonance energy is small compared to the real part $E$.
Unlike common discussions of complex resonance states the above derivation is based on the intuitive picture of a matter wave flowing out of a potential well, emphasizing the role of the conservation
of the probability current expressed by the continuity equation.


The above treatment for symmetric potentials can be straightforwardly generalized to resonances of asymmetric barrier potentials where the matter wave is localized between two maxima at $x=b_-<0$ and $x=b_+>0$ with two finite ranges 
$a_-,b_-$ and $a_+,b_+$ but then the actual calculations in step (\ref{I1}) are generally more difficult because the symmetry condition (\ref{BC_Sym}) no longer applies. Now one has to consider the problem of transmission through the barrier
and calculate the states $\psi_{\rm approx}$ and the respective real energies $E_{\rm approx}$ which correspond to local maximima of the transmission coefficient with $|T(E_{\rm approx})|^2<1$. 
In analogy to the symmetric case one finds that the relation (\ref{Gamma_s}) for the decay 
coefficient is generalized to
\be
   \Gamma_{\rm approx}=\frac{\hbar^2 k_{\rm approx} |\psi_{\rm approx}(a_+)|^2+\hbar^2 k_{-a} |\psi_{\rm approx}(a_-)|^2}{m \int_{b_-}^{b_+} |\psi_{\rm approx}(x)|^2 \rd x}
   \label{Gamma_asym}
\ee
where $k_{approx}=\sqrt{2mE_{\rm approx}}/\hbar$.
An important special case of an asymmetric barrier is a trap which is open on one side only whereas there is an impenetrable barrier on the other side. If the impenetrable barrier is on the 
left hand side (at $x=a_-$) we obtain the boundary condition
\be 
   |\psi_{\rm approx}(a_-)|^2=0
\label{dings}
\ee
and the wavefunction $\psi_{\rm approx}$ and the corresponding energy $E_{\rm approx}$ can be obtained by finding an approximate solution to the system of equations given by (\ref{dings}) and 
\be
  \psi_{\rm approx}'(a_+)= \ri k_{\rm approx} \psi_{\rm approx}'(a_+) \,.
  \label{Gamma_oneside}
\ee
An example of such a calculation is given in section \ref{subsec:delta_shell}. 
Note that for such a potential exact solutions for real energies do not exist.  
We further note that potentials with infinite range can usually be approximated by finite range potentials by choosing appropriate values for $a_\pm$ and that the Siegert approximation method 
can be generalized to two and three dimensions (see \cite{08nlLorentz} for a detailed discussion).

\section{Applications}
\label{sec:apps}
\subsection{The finite square well potential}
\label{subsec:square_well}

\begin{figure}[htb]
\begin{center}
\includegraphics[width=0.3\textwidth] {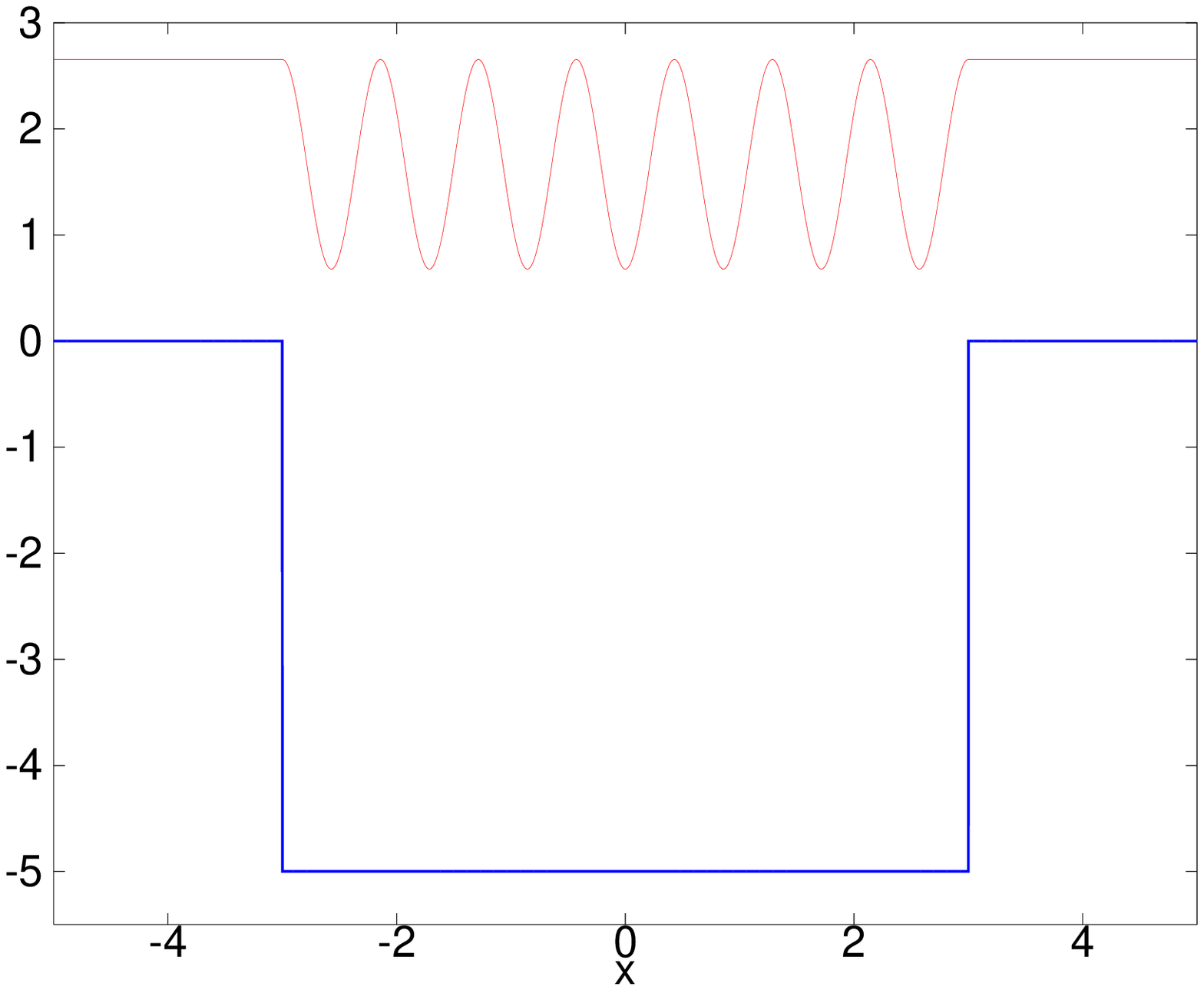}
\includegraphics[width=0.3\textwidth] {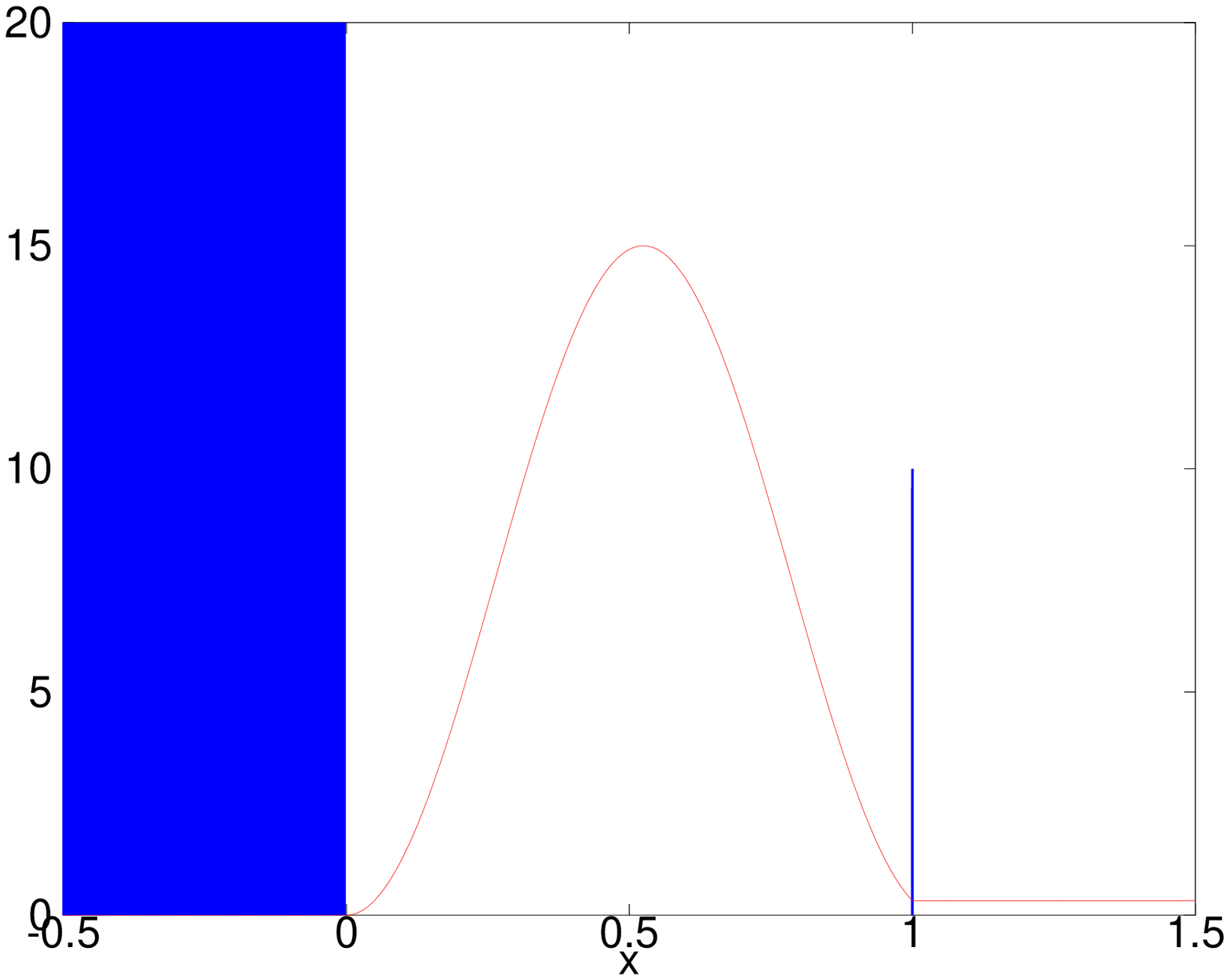}
\includegraphics[width=0.3\textwidth] {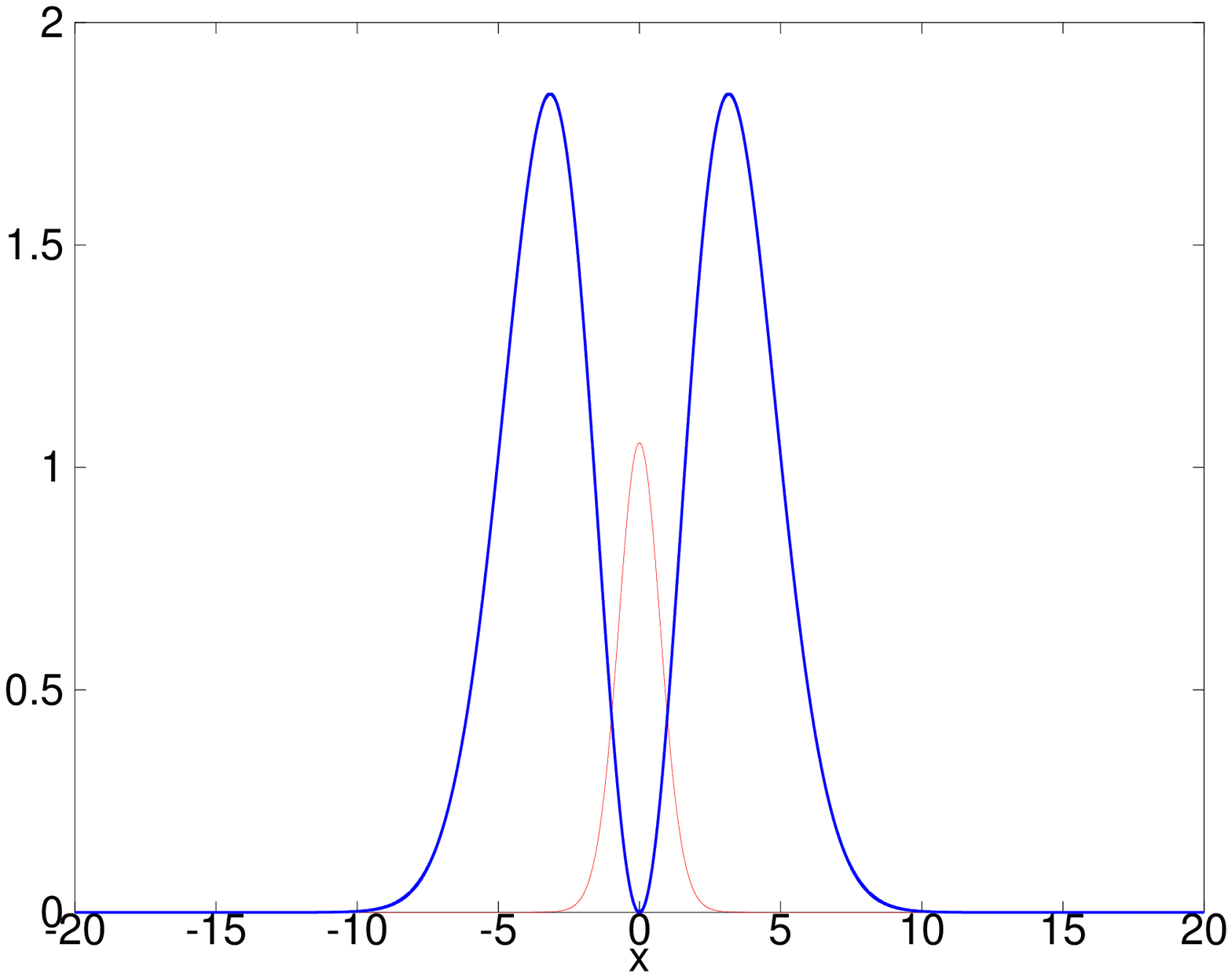}
\caption{\label{fig:Siegert} Barrier potentials (bold blue) and approximate squared wavefunctions $|\psi|^2$ (red) of the corresponding most stable resonance states (units with $\hbar=m=1$). 
Left: Finite square well (\ref{square}) with $V_0=-5$, $L=3$.
Middle: Delta Shell potential (\ref{Shell}) with $\lambda=10$, $L=1$ (The Dirac delta function is symbolically represented by a vertical line). 
Right: Double barrier potential (\ref{double}) with $V_0=1$, $\lambda=0.1$.}
\end{center}
\end{figure}

As a first analytically solvable example we consider the finite square well potential 
\be
   V(x)=\left\{
                \begin{array}{cl}
		0 & |x|>L\\
		V_0<0 & |x| \le L
		\end{array}  \right. 
\label{square}
\ee
with width $2L$ which is shown in the left panel of figure \ref{fig:Siegert}. As discussed at the beginning of the preceding section, the complex resonances for such a symmetric potential can be calculated approximately by finding its transmission 
resonances which satisfy the symmetry and boundary conditions (\ref{BC_Sym}) with $\alpha=T$. With the notation of the preceding section we can identify $b=a=L$, i.e. both the position of the local 
maximim and the range of the potential are given by the box length parameter $L$. Dropping the index $T$ we can write
the transmission resonance wave function inside the square well as a superposition of plane waves,
\be
   \psi(x)=F \re^{\ri q x} + G \re^{-\ri q x} \, , \quad |x|<L
   \label{f1}
\ee
where $q=\sqrt{2m(E-V_0)}/\hbar$ is real. The symmetry condition $|\psi(-x)|^2=|\psi(x)|^2$ leads to $FG^*=GF^*$ so that
\be
   |\psi(x)|^2=|F|^2+|G|^2+2FG^*\cos(2qx) \, .
   \label{fq1}
\ee
The second condition in (\ref{BC_Sym}) reads $\psi'(L)=\ri k \psi(L)$. Inserting Eq.~(\ref{f1}) we obtain
\be
   G=\frac{q-k}{q+k} \re^{2 \ri qL} F \, .
   \label{G1}
\ee
The condition $FG^*=GF^*$ then implies $\exp( 4 \ri qL)=1$ which leads to the resonance condition $2qL=n \pi$ with integer $n$. Thus we arrive at the celebrated formula
\be
   E_n=V_0+ \frac{\hbar^2 q^2}{2m}=V_0+ \frac{\hbar^2 \pi^2}{8mL^2}n^2 ,
   \label{ERlin}
\ee
for the transmission resonance energies of a finite square well potential. The number $n$ must be sufficiently large to make $E_n$ positive. Inserting Eq.~ (\ref{G1}) into (\ref{fq1}) leads to
\be
   |\psi(x)|^2=|F|^2 \left[1 +2(-1)^n\frac{q-k}{q+k}\cos(2qx) + \left(\frac{q-k}{q+k}\right)^2 \right]
\ee
( cf.~left panel of figure \ref{fig:Siegert}) and in particular
\be
   |\psi(L)|^2=|F|^2 \left[1+\frac{q-k}{q+k} \right]^2=|F|^2 \left(\frac{2q}{q+k} \right)^2 \, .
   \label{fa1}
\ee
Inserting $|\psi(L)|^2$ and the integral
\begin{eqnarray}
  \int_{0}^L |\psi(x)|^2 \rd x &=& L|F|^2 \left( 1+ \left(\frac{q-k}{q+k}\right)^2\right) \nonumber \\ 
                            &=& L|F|^2 \frac{(q+k)^2+(q-k)^2}{(q+k)^2} \nonumber \\
                            &=& 2L|F|^2 \frac{q^2+k^2}{(q+k)^2}
\end{eqnarray}
into the Siegert formula (\ref{Gamma_s}) we obtain the decay coefficient
\be
   \Gamma_n =\frac{2 \hbar}{L} \sqrt{ \frac{2 E_n}{m} } \frac{q^2}{q^2+k^2}
\ee
or
\be
   \Gamma_n=\frac{2 \hbar}{L} \sqrt{ \frac{2}{m} } \frac{\sqrt{E_n}(E_n-V_0)}{2E_n-V_0} \,
   \label{Gamma1}
\ee
which is the well known textbook result for the decay coefficient (or resonance width) of a finite square well potential (see e.g. \cite{Mess91}), that is usually obtained by expanding the transmission coefficient 
in a Taylor series around $E=E_n$.

\subsection{Delta Shell potential}
\label{subsec:delta_shell}
As an analytically solvable example for complex resonances in asymmetric potentials we consider the one-dimensional delta-shell potential

\be
 V(x)= \left\{
                    \begin{array}{cl}
                     +\infty   &    x \le 0 \\
                     (\hbar^2/m) \lambda \, \delta(x-L)     &   x>0\\ 
                     \end{array}
              \right. \, \text{ with } \lambda, L>0 \nonumber
\label{Shell}
\ee
consisting of an infinitely high potential barrier at $x=0$ and a delta barrier at $x=L$ which is illustrated in the middle panel of figure \ref{fig:Siegert}.
To find the solution of the corresponding Schr\"odinger equation 
\begin{equation}
 \left[ -\frac{\hbar^2}{2m} \frac{\rd^2}{\rd x^2} + V(x) \right] \psi(x) =
\left(E-\ri \Gamma/2 \right) \psi(x)
\end{equation}
we make the ansatz
\be
 \psi(x)= \left\{
                    \begin{array}{cl}
                      I\, \sin(k x)   & 0 \le x \le L \\
                     C \, \re^{{\rm i}kx}                              &     x>L 
                    \end{array}
              \right.  \, \text{ and }  E- \ri \Gamma/2 =\frac{\hbar^2 k^2}{2m}   \nonumber
\label{DShell_ansatz}
\ee
for the wavefunction which satisfies the required outgoing wave (Siegert) boundary condition for $x \rightarrow \infty$.
The matching conditions for the wavefunction at $x=L$ read
\be \psi(L-\epsilon)=\psi(L+\epsilon)\,, \quad \quad  \psi'(L-\epsilon)+2 \lambda \psi(L)=\psi'(L+\epsilon) \ee
where the discontinuity in the derivative is caused by the delta function potential \cite{Mess91}.
This leads to
\be I \sin(k L)=C \, \re^{{\rm i}kL} \, , \quad \quad k I \cos(kL)+2 \lambda I \sin(k L)= i k C \, \re^{{\rm i}kL} \ee
or 
\be 
   k \cos(kL)+(2 \lambda -\ri k)\sin(k L)= 0 \,.
\label{Dlin}
\ee
The real and imaginary part of the eigenenergy can now be found by numerically solving the transcendental equation (\ref{Dlin}) in the complex plane.
In the following we show how the Siegert approximation method presented in section \ref{sec:Sieg} can be used to obtain a convenient approximation in an analytically closed form. 
To obtain the approximate real part of the energy and approximate wavefunction as required in step (\ref{I1}) of the method we make the following approximation: 
Imagine that the delta potential at $x=L$ is infinitely strong, i.e. the limit $\lambda \rightarrow \infty$. Then the system is a closed box of length $L$ and the wavenumber satisfies $k L= n \pi$ with an integer $n$.
For a strong but still finite delta potential, i.e. $|k|<<\lambda$, we therefore assume $k L= n \pi + \delta L$ with $\delta L <0$ and $|\delta L| \ll 1$.
Inserting this ansatz into the real part of Eq.~(\ref{Dlin}) and expanding it up to second order in $\delta L$ yields
\be
   \delta =\frac{2 \lambda L +1}{n \pi L}-\sqrt{\left(\frac{2 \lambda L +1}{ n \pi L}\right)^2+\frac{2}{L^2}} \, .
\ee
Inserting $k= n \pi/L +\delta$ into equation (\ref{DShell_ansatz}) yields approximations for the resonance wavefunction and the real part of the eigenenergy. An approximation of the imaginary part
is obtained by inserting these results into the Siegert formula (\ref{Gamma_asym}),
\be
\fl \quad \quad \quad   \Gamma/2 \approx \frac{\hbar^2k |\psi(L)|^2}{2m\int_0^L |\psi(x)|^2 \rd x}=\frac{2 \hbar^2}{m} \left(\frac{n \pi}{L}+\delta \right)^2\frac{\sin^2(\delta L)}{n \pi +\delta L-\sin(2 \delta L)},
\ee 
where we have identified $b_+=a_+=L$ and $b_-=a_-=0$.
For a potential with $\lambda=10$, $L=1$ and scaled units $\hbar=m=1$ the Siegert approximation yields ${\cal E}_{approx}=4.481-\ri 0.062$ for the most stable resonance (cf.~middle panel of figure 
\ref{fig:Siegert}) which is in good agreement with the numerically exact result ${\cal E}_{exact}=4.487-\ri 0.061$.
A treatment of the equivalent problem within the context of the nonlinear Schr\"odinger equation can be found in \cite{09ddshell}.

\subsection{Double barrier}
\label{subsec:DGauss}
As an example of a numerical problem we consider the double barrier potential
\be
    V(x)=\frac{V_0}{2} x^2 \exp(-\lambda x^2) 
\label{double}
\ee
with $V_0=1$ and $\lambda=0.1$ so that the position of the potential maxima is given by $b = 1/\sqrt{\lambda} \approx 3.16$  using units where $\hbar=m=1$.
In order to calculate the approximate resonance wavefunction and real part of the energy as required in step (\ref{I1}) of the method we solve the boundary value problem given by equation (\ref{BC_Sym}) by means of a shooting procedure. 
We choose the cutoff parameter for the potential $a=20$ to ensure that $V(x) \approx 0$ for $|x|>a$ so that the wavefunction is well approximated by a plane wave in that region. Starting with initial conditions given in equation 
(\ref{BC_Sym}) we integrate the Schr\"odinger equation from $x=-a$ to $x=0$ using a standard Runge Kutta solver. By means of a bisection method the real energy $E$ is adapted such that the boundary 
condition at $x=0$ is satisfied. The decay rate is again obtained by inserting this value of $E$ and the corresponding wavefunction into the Siegert formula (\ref{Gamma_s}). More details on the actual numerical
implementation of the method can be found in \ref{app:num}.

\begin{table}[htbp]
\caption{\label{tab-compare-Sieg-cs} Complex eigenenergies for the three most stable resonances of the potential (\ref{double}) calculated using complex scaling (CS) and the Siegert approximation (S).} 
\begin{indented}
\item[] \begin{tabular}{lll}
\br
 $n$ & $\cal{E}_{\rm S}$ &  $\cal{E}_{\rm CS}$   \\      
\mr
  1    & $0.4601- \ri \, 9.62  \times \, 10^{-7}$  & $0.4601 -  \ri \, 9.6204 \times \, 10^{-7}$  \\ 
  2    &  $1.2804 -\ri \, 1.70 \times \, 10^{-3}$  & $1.2804 - \ri \, 1.6737 \times \, 10^{-3}$  \\ 
  3    &  $1.88 {\lineup \0\0}- \ri \, 7 {\lineup \0} \times \, 10^{-2}$    & $1.8531 -  \ri \, 6.7240 \times \, 10^{-2}$  \\ 
\br
  \end{tabular}
\end{indented}
\end{table}

The right panel of figure \ref{fig:Siegert} shows the potential (\ref{double}) and the square $|\psi|^2$ of the most stable resonance wavefunction calculated with the Siegert approximation method.   
Table \ref{tab-compare-Sieg-cs} compares our results for the three most stable resonances with numerically exact results calculated with the complex scaling method described in \cite{02computing}. 
We see that our simple approximation yields very good results for the ground state since its decay rate is small. The values for the first and second excited states demonstrate that the Siegert
 approximation becomes less accurate with increasing decay rates. A generalization of the same problem to the nonlinear Schr\"odinger equation is straightforward and can be found in 
\cite{08nlLorentz}.

\section{Summary and conclusion}
\label{sec:summary}

In this article the Siegert approximation method for calculating complex resonance states was presented in an intuitive and straightforward manner which at the same time clearly points out the similarities
and differences between resonances of transmission coefficients (or similar quantities) and resonances in the complex plane (Siegert rsonances) as well as the role of the continuity equation in this context. 
It was illustrated by two analytically solvable example problems and a numerical application. The author hopes that the present article, in addition to drawing attention to a useful and easily applicable 
computational tool, offers an alternative, rather intuitive point of view for a better understanding of complex resonances which complements other, more technical treatments.

\ack
 The author would like to thank Hans J\"urgen Korsch and Nina Lorenz for helpful discussions and comments. Financial support from a scholarship of the Universit\'e Libre de Bruxelles is gratefully acknowledged.

\begin{appendix}
\section{Numerical calculation}
\label{app:num}

The following commented MATLAB code implements the numerical algorithm for calculating resonances of the one-dimensional symmetric finite range potential described in section \ref{subsec:DGauss}.
For paedagogical reasons the program makes use of several global variables which give the code a simple structure but make it less flexible and elegant. For the same reasons and for achieving 
compatbility with both MATLAB and the Open Source software OCTAVE the integration of the Schr\"odinger equation is performed by means of a straightforward implementation of the classical Runge 
Kutta method which requires a rather high number of grid points. Thus the present code can be made a lot more efficient by using more sophisticated integrators like, e.g., MATLAB's 
{\verb ODE45 }  or OCTAVE's {\verb lsode }. The program can be straighforwardly adapted for other symmetric potentials by changing the function $V(x)$ and providing the corresponding 
position $b$ of the potential maximum that confines the wavefunction as well as a suitable cutoff parameter $a$. For a potential open on one side only the boundary condition criterion $crit$ 
must be modified according to equation (\ref{dings}).

\begin{lstlisting}[language=Matlab, breaklines=true]
 
function [E_res, Gamma, psi, Dpsi, x]= Num_Siegert  
% Calculates complex resonances of a double barrier potential
% with the Siegert approximation method
global V0 lambda m hbar a Nx

% Parameters for the double barrier potential: 
V0=1; lambda=0.1;

% hbar and mass:
hbar=1; m=1;

% Shooting method for calculating the resonance wavefunction 
% and real part of the energy:

a=-20; % left boundary of integration interval [a, 0]
Nx=10000; % number of lattice points / steps for integration

Emin=0.3; % lower boundary for the energy
Emax=0.6; % upper boundary for the energy

% use bisection method to find the real part of the resonance 
% energy of the solution of the Schoedinger equation in the 
% energy interval [Emin, Emax] :

E_res=bisect(@solve_SE,Emin,Emax,10^-5)

% Calculate wavefunction 'psi' and its derivative 'Dpsi' 
%for the resonance energy 'E_res' ('x' is a vector of 
% grid points, 'crit' is (close to) zero in case of 
% resonance (see below) ):

[crit,psi,Dpsi,x]=solve_SE(E_res);
                                                                                                 
k=sqrt(2*m*E_res)/hbar; 
dx=(0-a)/Nx;
b=1./sqrt(lambda); % calculate position of potential maxima

% Use Siegert formula to calculate the decay rate 
% (cf. equation (18) in the text) :
Gamma=hbar^2*k/m*abs(psi(1)).^2./(sum(abs(psi).^2.*(x>=-b)*dx))                                                               
Gamma_half=Gamma/2

% plot solution:

% normalize wavefunction for plot:
psi=psi./sqrt(sum(abs(psi).^2.*(x>=-b).*dx));

figure(1);
hold on
plot(x,abs(psi).^2)   % plot wavefunction
plot(-x,abs(psi).^2)

plot(x,V(x),'r')   % plot potential
plot(-x,V(x),'r')
hold off
xlabel('x')


%------------------------------------------------------------
function [crit,psi_v,Dpsi_v,x_v]=solve_SE(E) 
% integrate the Scroedinger equation from x=a to x=0 
% for a given energy E;  Return values:
%  crit: crit=0 if boundary condition at x=0 is satisfied 
%   (cf. equation (11) in the text) 
%  psi_v: vector containing the wavefunction
%  Dpsi_v: vector containing the derivative of the wavefunction
%  x_v: vector containing the grid points of the integration
global m hbar a Nx

psi_v=[];    % empty vector for the wavefunction
Dpsi_v=[];   % empty vector for the derivative of the wavefunction
x_v=[];      % empty vector for the grid points

k=sqrt(2*m*E)/hbar; 

% Siegert boundary conditions at x=a:
x=a; psi=0.001;  Dpsi=i*k*psi;  

x_v=[x_v x];                
psi_v=[psi_v psi];
Dpsi_v=[Dpsi_v Dpsi];

dx=(0-a)/Nx; % Step size for integration

for j=1:Nx % integrate over Nx steps 
  y0=[psi; Dpsi];

  % Runge Kutta integration step:

  dy_dx0=Schroedinger(y0,x,E); % call to function 
       %that implements the Schroedinger equation

  yA=y0+0.5*dx*dy_dx0;

  dy_dxA=Schroedinger(yA,x+0.5*dx,E);
  yB=y0+0.5*dx*dy_dxA;

  dy_dxB=Schroedinger(yB,x+0.5*dx,E);
  yC=y0+dx*dy_dxB;

  dy_dxC=Schroedinger(yC,x+dx,E);

  y1=y0+dx*(dy_dx0+2*(dy_dxA+dy_dxB)+dy_dxC)/6;

  % Store result of integration step in vectors :
    psi=y1(1); Dpsi=y1(2);
    x=a+j*dx;  % increase x
    x_v=[x_v x];
    psi_v=[psi_v psi];
    Dpsi_v=[Dpsi_v Dpsi];
end;

% calculate criterion for the boundary condition 
% at x=0 (cf. equation (11) in the text):
crit=(psi*Dpsi'+psi'*Dpsi);                            
%------------------------------------------------------------
function dy_dx=Schroedinger(y,x,E) 
% implementation of the Schroedinger equation as a 
% system of two first order differential equations
global m hbar

dy_dx = [y(2); 
        2*m/hbar^2 * (V(x)-E).*y(1)];
%------------------------------------------------------------
function potential = V(x) % Double barrier potential
global V0 lambda

potential = V0.*0.5.*x.^2.*exp(-lambda.*x.^2);
%--------------------------------------------------------------
function [x0] = bisect(func,x1,x2,tol)  
% bisection method for solving a nonlinear equation

  f1=feval(func,x1); if f1 == 0, x0=x1; return; end
  f2=feval(func,x2); if f2 == 0, x0=x2; return; end
   if f1*f2 >= 0 | x1 >= x2
        x1, f1
        x2, f2
      error('No root found due to initial values x1, x2');
   end

  for i=1:100
    if x2 - x1 < tol return; end
    x0 = 0.5*(x2+x1);
    f0=feval(func,x0);
    if f0*f1 <= 0
      x2=x0; f2=f0;
    else
      x1=x0; f1=f0;
    end
  end
  error('No root found') 

\end{lstlisting}

\end{appendix}

\end{document}